\documentclass[12pt]{article}
\pdfoutput=1
\usepackage{amsmath,epsf,amssymb,latexsym,enumerate,cite,shadow,array,color}
\usepackage{slashed}
\usepackage{graphicx,wasysym,epsfig}

\DeclareFontFamily{U}{rsf}{} \DeclareFontShape{U}{rsf}{m}{n}{
  <5> <6> rsfs5 <7> <8> <9> rsfs7 <10-> rsfs10}{}
\DeclareMathAlphabet\Scr{U}{rsf}{m}{n} \makeatletter
\def\be{\begin{equation}}
\def\ee{\end{equation}}
\def\ba{\begin{array}}
\def\ea{\end{array}}

\newcommand{\beq}{\begin{equation}}
\newcommand{\eeq}[1]{\label{#1}\end{equation}}
\newcommand{\bea}{\begin{eqnarray}}
\newcommand{\eea}[1]{\label{#1}\end{eqnarray}}

\usepackage{ifpdf}
\ifx\pdfoutput\undefined
   \pdffalse
\else
   \pdfoutput=1
   \pdftrue
  \usepackage[pdftex]{hyperref}
\begin{document}

\begin{titlepage}

\hskip 1.5cm

\begin{center}

{\huge{Prediction of short time qubit readout via measurement of the next quantum jump
 of a coupled damped driven harmonic oscillator}}

\vskip 0.8cm  

{\bf \large Massimo Porrati$^{a}$ and Seth Putterman$^{b}$}  

\vskip 0.5cm

\noindent 
$^{a}$ CCPP, Department of Physics, NYU, 726 Broadway, New York NY 10003, USA \\
$^{b}$ Department of Physics and Astronomy University of California Los Angeles, CA 90095-1547 USA

\end{center}

\vskip 1 cm

\begin{abstract}
The dynamics of the next quantum jump for a qubit [two level system] coupled to a readout resonator [damped driven 
harmonic oscillator] is calculated. A quantum mechanical treatment of readout resonator reveals non exponential short time 
behavior which could facilitate detection of the state of the qubit faster than the resonator lifetime.

\end{abstract}

\vspace{24pt}
\end{titlepage}



A quantum system that is both driven and observed will execute deterministic changes in the amplitudes of occupation 
of its various levels $| i\rangle$ that are interrupted by quantum jumps. In order to measure the time $t_j$ 
between successive
quantum jump one has to be able to measure that a jump has not happened for times  $ t< t_j $. For an atom driven to 
fluorescence by a laser a measurement of the next quantum jump implies the ability to determine that no fluorescent photon
 have been emitted during the interval between jumps. Such a measurement does not imply that the quantum system is 
 closed and unitary during this interval. Instead, the ability to make such a null measurement makes a small irreversible, 
 though phase coherent change, in the wave function. This weak, non projective measurement dramatically changes 
 the temporal dynamics of the quantum system. In particular one can observe long periods of intermittency~\cite{1,2,3,4}
 where the phase coherence enables the reversal of a quantum jump before it occurs~\cite{5}. 

	Recently the next quantum jump has been measured for a few level transmon qubit that is dispersively coupled to a 
read-out resonator. For the purpose of our analysis the transmon is to be thought of as an atom with discrete levels and the 
resonator to which it is coupled is approximated as a damped driven quantized harmonic oscillator. In our analysis the 
resonator is not treated as a classical measuring apparatus. Instead the resonator and atom are calculated as a coupled 
quantum system~\footnote{ Including the resonator in the quantum system has been used often in the 
theory of quantum measurement by photodetectors. While we are aware of applications to quadrature, homodyne and 
heterodyne measurements, we are not aware of it being applied to either next-photon detection or to the computation of
lifetimes of atomic states from first principles. A few references particularly relevant to our work are the reviews~\cite{6,6a} 
and refs.~\cite{6b,6c}.}.  
A key result of this approach is that for times short compared to the decay time $1\over \kappa$ of the 
cavity the probability $W(t)$  that a jump has not occurred for a time $t$ is not distributed as an exponential or Poisson 
distribution but as
\beq
W(t)\sim\exp(-{\bar{n}\kappa^3 t^3/12}),
\eeq{m1}
where $\bar{n}$ is a measure of the strength of the drive. ($\bar{n}$ is the occupation number that characterizes the 
steady state of the damped driven oscillator when treated as a classical system.) 
The probability that the next jump will take place in the interval $[t,t+dt]$ is:
\beq
D(t)= -{dW\over dt} dt .
\eeq{m2a}
In the limit of large $\bar{n}$~(\ref{m1},\ref{m2a}) imply that the next quantum jump can occur on a time scale shorter 
than the 
lifetime, $1/\kappa$, of the resonator. To the extent that the resonator is a proxy for the qubit, a measurement of the next 
jump yields information on the state of the qubit on this short timescale. 
We also calculate how the coupling to 
the resonator determines the effective lifetime of the excited states of the atom. We work in the limit of perfect
detection, in which the damping coefficient, $\kappa$, is due only to measurement. We also assume the validity of the
 dispersive approximation~\cite{7}.

The system is described by the amplitudes $C_{G,n}; C_{B,n}$ for the resonator to be in level $n$ and the 
atom to be in its ground state
 $G$ or excited state $B$, subject to the condition that there has been no quantum jump between a reset at $t=0$ and 
 time $t$. The probability that the next jump has not occurred during an interval of  time $`t'$ is then   
 \beq
W(t)=\sum_{n=0}^\infty \left[ |C_{G,n}(t)|^2+ |C_{B,n}	(t)|^2 \right] .
\eeq{m2}

The temporal response~\eqref{m1} already appears in the quantum dynamics for the next quantum jump of the driven 
damped resonator by itself. In this case we consider the atom in state $G$ and the resonator being driven at it resonant 
frequency $\omega_C$ when the atom is in its ground state. In the rotating wave approximation the equation of motion for 
the resonator amplitudes is:
\beq
{dC_{G,n}\over dt}=\Gamma [ \sqrt{n}C_{G,n-1} -\sqrt{n+1} C_{G,n}] -{\kappa n \over 2} C_{G,n},
\eeq{m4}
where the externally imposed drive is $\Gamma=\kappa \sqrt{\bar{n}}/2$. In the absence of damping, $\kappa$, the initial 
state $C_{G,n}(0)=1$ is lifted by $\Gamma$ to higher levels and the norm is preserved. But with damping the norm of the 
wave function --projected onto the space that has not jumped-- is decreasing. This is contained in the exact solution 
to~\eqref{m4}:     							
\bea
C_{G,n}(t) &=& e^{\beta(t)}{\alpha(t)^n \over \sqrt{n!}}, \label{m5} \\ 
\beta(t) &=& -{\kappa \over 2}\bar{n}\left[ t +{2\over \kappa} \left( e^{-\kappa t/2} -1\right)\right] ;  \label{m5a} \\
{d\beta \over dt} &=& -\Gamma \alpha(t),   \label{m6} \\
W(t) &=& \exp[\beta + \beta^* + |\alpha|^2 ] .
\eea{m7}								
This exact solution is the exponential of an exponential. In the limit of small times $W(t)$ is given by~\eqref{m1}, and the average time between the initial condition and the first quantum jump between some level of the resonator is: 
\beq
t_j =	-\int_0^\infty t 	{dW \over dt} dt		.			
\eeq{m8}
At short times $\bar{t}_j \sim a_0(3/\kappa \Gamma^2)^{1/3}$ where $a_0=\Gamma(1/3)/3$. If  $\bar{n} \gg 1$, 
$\kappa  \bar{t}_j<1$ and the dynamics of the next quantum jump is dominated by the non-exponential behavior 
in~\eqref{m1}. For $\kappa t >1$, $W \rightarrow W_0 \exp(-\kappa \bar{n}  t)$ which is exponentially distributed. 
	
If the drive frequency $\omega_D$ is detuned from the natural resonant frequency $\omega_C$ so that, 
$\omega_D = \omega_C-|\chi |$, then the RWA yields the equations of motion of the resonator when the atom is in $G$:
\beq
{dC_{G,n}	\over dt}=in|\chi |C_{G,n} + 	{\kappa \sqrt{\bar{n}} \over 2} [ \sqrt{n}C_{G,n-1} -\sqrt{n+1} C_{G,n}]	 
-{n\kappa \over 2}C_{G,n}	.
\eeq{m9}					           
The solution is given by~\eqref{m6} and:
\bea
\alpha &=& {i\Gamma \over i(\kappa /2) +\chi} \{ 1- \exp[-(\kappa/2)t+i\chi t ] \} , \label{m10} \\
{dW\over dt} &=& -\kappa \alpha \alpha^* W(t) . \nonumber  
\eea{m9a}
 If $\kappa/\chi \sim 1$ the detuning is irrelevant, so the new interesting limit is $\kappa /\chi \ll 1$. If in addition 
$\kappa t \gg 1$:
\beq
W(t)\sim \exp(- \Gamma^2/\chi^2 - \Gamma^2 \kappa t / \chi^2 ) ,
\eeq{m11}
which is the exponential in time decay typical of an excited quantum system with discrete levels. A large detuning
is one of the conditions under which a dispersive Hamiltonian approximates the photon-cavity dynamics~\cite{7} so this
is anyway the regime of greatest physical interest to us. For large dispersion and short time [ $\kappa t < 1 $]: 	 
\beq
W(t) \sim \exp\left[	-{\kappa^3 \bar{n} \over 2 \chi^2 }\left( t- {\sin \chi t \over \chi} \right)	\right] .
\eeq{m12}
When additionally $\chi t <1$  this again gives the distribution of dark times in Eq.~\eqref{m1}. 
The percentage of jumps to happen during the time when Eq.~\eqref{m1} applies is 
$$
{\kappa^3 \bar{n} \over 12 \chi^3}.
$$

When the resonator is coupled to an atom there is the possibility of Rabi flopping at a frequency $\Omega$  between the 
ground state $G$ and the excited state $B$. For the dispersively coupled system the resonators natural frequency depends
 upon the state of the atom. When the atom is in $G$ the resonant frequency of the oscillator is $\omega_C$ and when the 
 atom is in $B$ the resonant frequency is $\omega_C-|\chi |$ . Following the experimental arrangement~\cite{5} the external   
 drive is tuned to the upper level [or bright level] resonant frequency so that: $\omega_D=\omega_C-|\chi | $. In this case  
   the equations that give the amplitude for the next quantum jump are:
\bea
{dC_{B,n}\over dt} &=& i \Omega C_{G,n} + \Gamma [ \sqrt{n}C_{B,n-1} -\sqrt{n+1} C_{B,n}] -{n\kappa \over 2} C_{B,n}
\label{m13} \\
{dC_{G,n}\over dt} &=& i \Omega^* C_{B,n} + \left[{dC_{G,n} \over dt}\right]_R
\eea{m14}
where $[dC_{G,n}/dt]_R$ is given by the rhs of~\eqref{m9}. A closed form approximation of the solution to~\eqref{m14} can
be obtained via multi-time scale analysis in the limit of large dispersion. In the limit where $\Gamma^2/\chi^2 \ll 1$, 
$|C_{G,2}/C_{G,1}| \ll 1$  and we neglect $C_{G,n}$  for $n>1$. This leads to the next jump equations:
\bea
 {dC_{B,0} \over dt} &=& 	i\Omega	C_{G,0}	-\Gamma 	C_{B,1} , \label{m15} \\
 {dC_{G,0}\over dt} &=&	i \Omega^* C_{B,0}  -\Gamma 	C_{B,1} . \label{m16} \\
 {dC_{G,1}\over dt} &=& (i\chi -\kappa/2)C_{G,1} +\Gamma C_{G,0} .
 \eea{m17}
These equations can be closed by noting that:
\beq
C_{B,1}(t) = \int ds {dC_{B,0}\over ds} \alpha(t-s)\exp[\beta(t-s)].
\eeq{m18}
More generally in the multiscale approximation the occupation of the excited resonator levels $|B,n\rangle$ is driven by the occupation of $|B,0\rangle$:
\beq
|B(t)\rangle = \sum C_{B,n}(t) | B,n\rangle =\int ds {dC_{B,0}\over ds} \exp[\alpha(t-s)a^\dagger +\beta(t-s)] |0\rangle.
\eeq{m19}
Using \eqref{m5a} for  $\beta$ in combination with the saddle point method which can be applied when $\bar{n}\gg 1$, yields: $C_{B,1}=\sqrt{2/\pi} C_{B,0}$.  
The eigenvalues of (\ref{m15}-\ref{m17}) are determined by the solution to the cubic equation:		
\beq
\lambda(\lambda + \beta_B/2)(\lambda -i\chi +\kappa/2) 	+\Omega^2(\lambda -i\chi +\kappa/2) +\Gamma^2 
(\lambda +\beta_B/2)=0,
\eeq{m20}
where $\beta_B/2=\sqrt{2/\pi} \Gamma$. 
For large dispersion the next jump can come from either of the transitions $|G,1\rangle \rightarrow |G,0\rangle $ or 
$|B,1\rangle \rightarrow |B,0\rangle$. In order that the relative probability of a jump originating from the excited $B$ state be 
large and furthermore take place on a long time scale we must require that 
\beq
\beta^2_B/4 > \Omega^2 > \kappa \beta_B \Gamma^2 /4\chi^2.
\eeq{m21a}
In this limit the eigenvalues are approximately:
\beq
\lambda_1=-\beta_B/2; \qquad \lambda_2=-\gamma; \qquad \lambda_3=i\chi -\kappa/2
\eeq{m22a}
where  $\gamma=2|\Omega|^2/\beta_B$.
 The coefficient that controls the effective lifetime of the 
  upper level of an ideal qubit is determined by the coupling to the resonator drive.  The solution to~(\ref{m15},\ref{m20}) 
  displays motion on two well separated time scales: the short time scale $2/\beta_B$  and the long time scale 
  $1/\gamma$. If the next jump has not occurred on the short time scale then there is a lull and a 
  warning that the jump will take place on the long time scale $1/\gamma$. The time to the next jump for the 2 level system coupled to 
  the cavity depends on the norm of the system \eqref{m2}: 
\beq
  W(t)= \langle 	B(t) | B(t) \rangle +|C_G(t)|^2 .
\eeq{m20a}
For large $\bar{n}$ there is again a regime of time-scales with non-exponential behavior, this time for the 2 level system. 
 When $\bar{n}^{1/6}> \Gamma t > 1$, the rate of the next quantum jump becomes: 
\beq
 -{dW\over dt} 	= 2\gamma W	-2\gamma \exp\left[ -\bar{n}(\kappa t)^3/12\right] .
 \eeq{m21}
 
 \begin{figure}[h]
\begin{center}
\epsfig{file=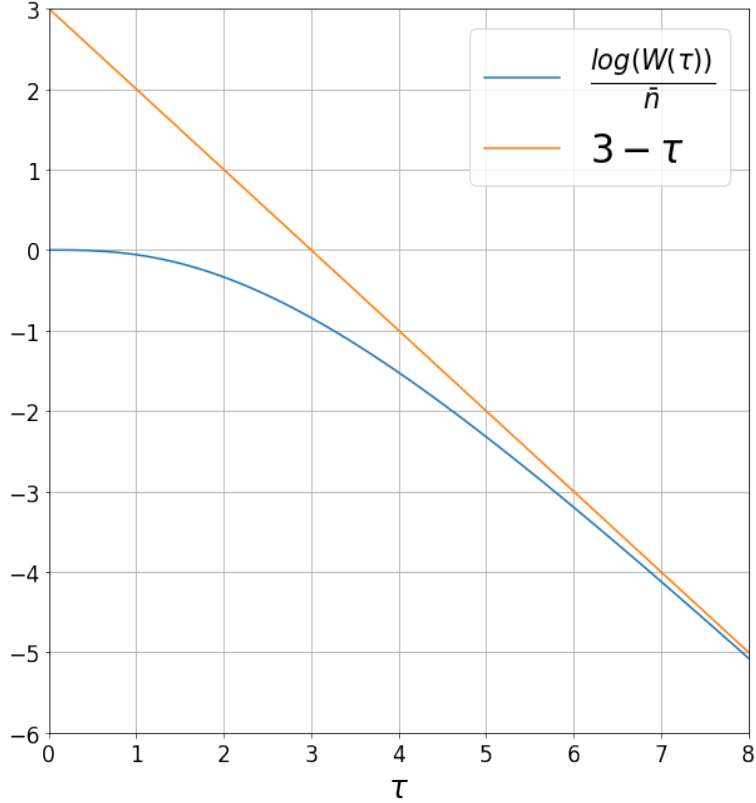, height=5.5in, width=4in}
\end{center}
\caption{The graph of the logarithmic norm of the system vs $\tau=\kappa t$:
$\log[W(\tau)]/\bar{n}=-\tau + 2\alpha(\tau)/\sqrt{\bar{n}} +\alpha^2(\tau)/\bar{n}$. It shows that the time evolution changes 
from $\tau^3$ to $3-\tau$ at times $t=O(1/\kappa)$. A precise determination of the short-time behavior of the norm and the 
next quantum jump rate is essential for qubit readouts at times $t\lesssim 1/\kappa$.}
\label{figure1}
\end{figure}

Exponential time evolution of continuously observed systems appeared in previous work on quantum measurement 
theory and experiment of cavity-atom systems, see especially~\cite{7a,8}. We are not aware of previous works exhibiting 
time evolution $O[\exp(t^3)]$ as in our eq.~\eqref{m21}.
We propose that this short time behavior appears because we evaluate the amplitude for the next jump as compared to 
taking an average over the occupation levels of the resonator/cavity.

\begin{figure}[h]
\begin{center}
\epsfig{file=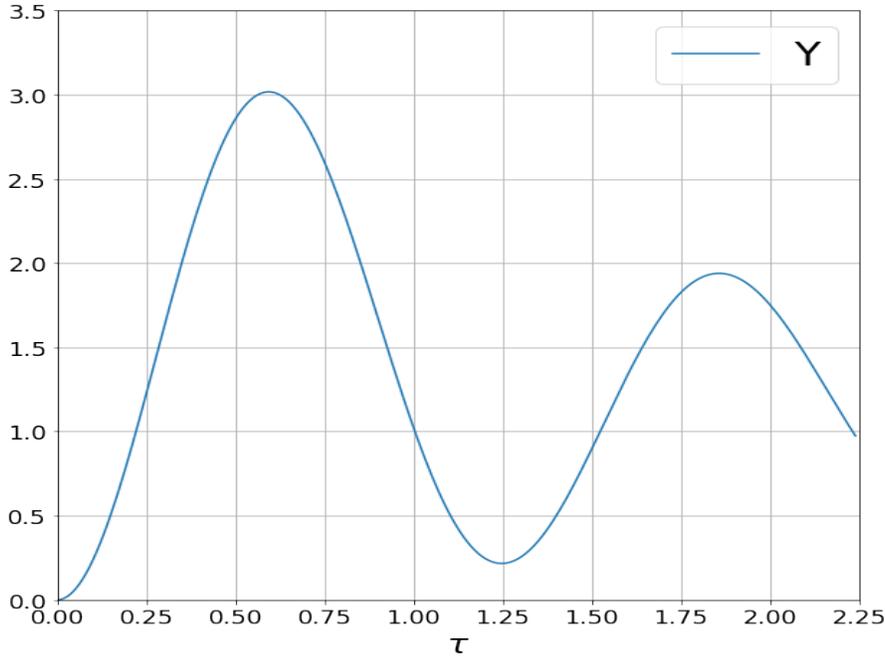, height=3.5in, width=6in}
\end{center}
\caption{The scaled logarithmic decrement $Y$ of probability for next quantum jump as a function of time.}
\label{figure2}
\end{figure}

For the 2 level atom dispersively coupled to a resonator a measurement of the next jump of the resonant cavity yields 
information on the state of the qubit. If the qubit is in $|B\rangle$ and the drive frequency is on resonance for this state then
the expected time to the next jump is $\bar{t}_j \sim (12/\bar{n})^{1/3}/\kappa$. For sufficiently large $\bar{n}$ this time is
shorter than $1/\kappa$. 
The norm $W$ for the state which has not yet jumped is 
 plotted in Figure~(\ref{figure1}).
 If instead of being in $|B\rangle$ the qubit was in state $|G\rangle$ when the drive is turned on there is also a chance of 
recording a jump. In the limit of large dispersion this ``error'' rate, $\epsilon$ is approximately the ratio of the change in the 
norms for the different initial conditions:
\beq
\epsilon \sim [1- W(\chi,t_j)]/[1-W(0,t_j)] .
\eeq{m22}
For $\kappa t_j < 1$, $\epsilon \sim 2\kappa t_j\Gamma^2 /\chi^2 \sim [\kappa\bar{n}^{1/3}/\chi]^2$ where dispersion is
sufficiently large that $\epsilon \ll 1$. For large dispersion the time dependence of the probability for next photon emission 
goes through large oscillations. Figure~(\ref{figure2}) displays the scaled logarithmic decrement $Y$ of the norm as a 
function of time for $\chi/\kappa=5$:
 \beq
 Y= -{1+(2\chi/\kappa)^2 \over \bar{n} W} {dW\over d\tau} ={1+(2\chi/\kappa)^2 \over \bar{n} } \alpha\alpha^*
 =\left[1- e^{-\tau/2} \right]^2 + 4 e^{-\tau/2} \sin^2(\chi\tau/2\kappa).
 \eeq{m23}
 
The time scale for determining the state of the qubit is shorter than the cavity lifetime when $\bar{n}>12$. 
 ``Reducing the time required to distinguish qubit states with high fidelity is a critical goal in quantum-information 
 science~\cite{9}.'' 
 So we propose considering whether a method based upon the next jump of the quantum states of the resonator will yield a more effective determination of the state of the qubit, than methods which take longer than $1/ \kappa$~\cite{5}. We further remark that the resonator lifetime $1/\kappa$ differs from spontaneous decay in that it is not a first principles quantity. 
 The resonator decay is an ensemble average of thermodynamic/ transport properties of the cavity. 
 The irreversible processes that determine $1/\kappa$ take place on much shorter timescales. This suggests the possibility 
 of varying $\kappa$ on a time scale long compared to these micro-processes, but short compared to the long time scale 
 of a quantum jump, $1/\gamma$ . We suggest that such changes will affect the phase of the between-jumps 
 wave function while leaving the time evolution coherent.


 \subsection*{Acknowledgments} 
 We wish to acknowledge valuable discussions with Hong Wen Jiang.

\end{document}